\documentclass[twocolumn,showpacs,preprintnumbers,
amsmath,amssymb,aps,prd,nofootinbib,superscriptaddress,
eqsecnum]{revtex4}
\usepackage{graphicx,color}

\makeatletter
\renewcommand{\p@subsection}{}
\makeatother

\newcommand{\Slash}[1]{\ooalign{\hfil/\hfil\crcr$#1$}}

\begin{document}

\title{
Density fluctuations
in the presence of spinodal instabilities}

\author{C. Sasaki}
\affiliation{%
Gesellschaft f\"ur Schwerionenforschung, GSI,  D-64291 Darmstadt,
Germany}
\author{B. Friman}
\affiliation{%
Gesellschaft f\"ur Schwerionenforschung, GSI,  D-64291 Darmstadt,
Germany}
\author{K. Redlich}
\affiliation{%
Institute of Theoretical Physics, University of Wroclaw, PL--50204
Wroc\l aw, Poland}
\affiliation{%
Institute f\"ur Kernphysik, Technische Universit\"at Darmstadt,
D-64289 Darmstadt, Germany}

\date{\today}

\begin{abstract}
Density fluctuations resulting from spinodal decomposition in a
non-equilibrium first-order chiral phase transition are explored.
We show that such instabilities generate divergent fluctuations of
conserved charges along the isothermal spinodal lines appearing in
the coexistence region. Thus, divergent density fluctuations could
be a signal not only for the critical end point but also for the
first order phase transition expected in strongly interacting
matter. We also compute the mean-field critical exponent at the
spinodal lines. Our
analysis is performed in the mean-field approximation to the NJL
model formulated at finite temperature and density. However, our
main conclusions are expected to be generic and model independent.
\end{abstract}


\setcounter{footnote}{0}

\maketitle


One of the central questions addressed in the context of QCD is the
phase structure and the phase diagram of strongly interacting
matter at finite temperature and baryon number density \cite{pd}.
Based on calculations in effective models and on universality
arguments one finds that the order of the transition from the
hadronic to the quark gluon plasma phase depends on the number of
quark flavors and on the value of the quark
masses~\cite{pisarski,ef5,ef1,Klevansky,ef2,ef3,our,hatta,fujii}.
For physical values of the parameters one expects that the
transition at high temperature and low net baryon number density is
continuous. In the opposite limit of low temperature and large
density the QCD phase transition is expected to be first order.
This suggests that the phase diagram exhibits a critical end point
(CEP), where the first order chiral transition of QCD terminates
\cite{ef5,stephanov}. Recent, however still preliminary, results
obtained in first principle calculations of QCD, in lattice gauge
theory (LGT), confirm the existence of such a point in the
temperature T and chemical potential $\mu_q$ plane
\cite{LGT1,LGT3,LGT2}.

The search for the CEP has recently attracted considerable
attention. It is of particular interest to identify the position of
the critical end point in the phase diagram and to study generic
properties of thermodynamic quantities in its vicinity.  The
qualitative behavior of physical observables and their dependence
on thermodynamic variables in the critical region can be studied in
effective chiral models. However, a quantitative description of the
thermodynamics near the phase transition can only be obtained
theoretically by solving QCD {\em ab initio} in LGT or
phenomenologically in the context of experimental studies of heavy
ion collisions.

To locate the phase transition line in the QCD phase diagram one
needs observables that are sensitive probes of the critical
structure. Modifications in the magnitude of fluctuations or the
corresponding susceptibilities  have been suggested as a possible
signal for deconfinement and chiral symmetry restoration
\cite{hatta,stephanov}. In this context, fluctuations related to
conserved charges are of particular interest \cite{fluct}. The
fluctuations of baryon number and electric charge diverge at the
critical end point while they are finite along the cross over and
first order phase boundaries \cite{hatta,our,bj}. Consequently,
singular fluctuations of baryon number and electric charge as well
as a non monotonic behavior of these fluctuations as functions of
the collision energy in heavy ion collisions have been proposed as
possible signals for the QCD critical end point
\cite{hatta,stephanov, our}. However, the finiteness of the
fluctuations along the first order transition depend on the
assumption that this transition appears in equilibrium.

A first order phase transition is intimately linked with the
existence of a convex anomaly in the thermodynamic pressure
\cite{ran}, which can be uncovered only in non-equilibrium systems.
There is an interval of energy density or baryon number density
where the derivative of the pressure is positive, $\partial
P/{\partial V}>0$. This anomalous behavior characterizes a region
of instability in the ($T,n_q)$-plane, where $n_q$ is the net quark
number density. This region is bounded by the spinodal lines, where
the pressure derivative with respect to volume vanishes. The
derivative taken at constant temperature and that taken at constant
entropy define the isothermal and isentropic spinodal lines,
respectively.

The consequences of spinodal decomposition have been discussed in
connection with the chiral/deconfinement phase transition in heavy
ion collisions \cite{ran,gavin,rans,ranhi,heiselberg,polony}.
Furthermore, spinodal decomposition plays a crucial role in the
description of the 1st order nuclear liquid-gas transition in low
energy nuclear collisions \cite{ran,heiselberg}. It has also been
argued that in the region of phase coexistence, a phase separation
can lead to an enhancement of baryon \cite{gavin} and strangeness
fluctuations \cite{rans}.

In this letter we consider the fluctuations of conserved charges
along the spinodal lines, expected at finite net baryon density in
the QCD phase diagram. We show that if the chiral phase transition
is first order, then the fluctuations of the net densities of
baryon number and electric charge diverge along the isothermal
spinodal lines. Consequently, large fluctuations of these
quantities may be a signal for a first order phase transition in
the QCD medium. We also compute the mean-field critical exponent of
the quark susceptibility at the spinodal lines and compare with
that at the CEP.

 \setcounter{equation}{0}
%
In our study of fluctuations across the first order chiral phase
transition we adopt  the  Nambu--Jona-Lasinio (NJL) model. For two
quark flavors and three colors  the NJL Lagrangian reads
\cite{nambu,review}:
\begin{align}\label{eq1}
{\mathcal L} = \bar{\psi}( i\Slash{\partial} -m + \mu\gamma_0)\psi  
{}+ G_S \Bigl[ \bigl( \bar{\psi}\psi \bigl)^2 + \bigl( \bar{\psi}i\vec{\tau}\gamma_5\psi
\bigl)^2  \Bigr]\,,
\end{align}
where $m = \mbox{diag}(m_u, m_d)$ is  the current quark mass, $\mu = \mbox{diag} (\mu_u,
\mu_d)$ are the  quark chemical potentials  and $\vec{\tau}$ are Pauli matrices. The
strength of the interactions between the constituent quarks  is
controlled by the coupling $G_S \Lambda^2 = 2.44$ with the three
momentum cut-off $\Lambda=587.9$ MeV, introduced to regularize the
ultraviolet divergences. The parameters are fixed to reproduce the
pion decay constant in vacuum and the pion mass for $m_u=m_d=5.6$
MeV.

In the mean field approximation the thermodynamics of the NJL model
is, for an isospin symmetric system, given by the thermodynamic
potential \cite{review}:
\begin{align}\label{eq2}
& \Omega (T,\mu;M)/V = \frac{(M-m)^2}{4G_S} {}- 12 \int\frac{d^3p}{(2\pi)^3}
\Bigl[E(\vec{p}\,)
\nonumber\\
&
{}- T\ln ( 1-n^{(+)}(\vec{p},T,\mu)
)\Bigr.
{}-\Bigl.  T\ln (1-n^{(-)}(\vec{p},T,\mu) \Bigr]
\end{align}
where  $M = m- 2G_S\langle \bar{\psi}\psi \rangle$ is the dynamical
quark mass, $E(\vec{p}\,) = \sqrt{\vec{p}^{\,2} + M^2}$ is the
quasiparticle energy and
 $n^{(\pm)}(\vec{p},T,\mu) = \Bigl( 1 + \exp\bigl[ (E(\vec{p}\,)
\mp \mu)/T \bigr] \Bigr)^{-1}$ are the quark/antiquark
distribution functions.

The dynamical mass $M$ is obtained self consistently from the
stationarity condition ${\partial\Omega}/{\partial M} = 0$, which
implies
\begin{align}\label{eq3}
M = m+24 G_S \int\frac{d^3 p}{(2\pi)^3} \frac{M}{E} \Bigl[ 1 - n^{(+)} - n^{(-)}
\Bigr]\,.
\end{align}
The relevant thermodynamic observables, the quark number density
and the corresponding susceptibility, are given by
\begin{align}\label{eq4}
n_q &= -\frac{\partial \Omega}{\partial \mu}\,,
\qquad\qquad
\chi_{\mu\mu}=\frac{\partial n_q}{\partial\mu}\,.
\end{align}

%
\begin{figure}
\begin{center}
\includegraphics[width=8.6cm]{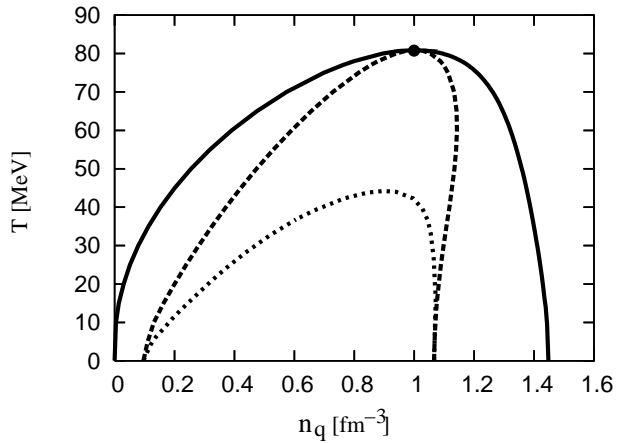}
\caption{  The phase diagram of the NJL model in the temperature
$(T,n_q)$-plane. The filled point indicates the CEP. The full lines
starting at the  CEP represent boundary of the coexistence region
in equilibrium. The broken-curves are the isothermal whereas the
dotted ones are the isentropic spinodal lines.
 }
\end{center}
\end{figure}
The NJL model has a generic, QCD like, phase diagram. It exhibits a
critical end point that separates the cross over from the first
order chiral phase transition. The relevant part of the phase
diagram in the $(T, n_q)$--plane is shown in Fig. 1. If the first
order phase transition takes place in equilibrium, there is a
coexistence region, which ends at the  critical end point. However,
in a non-equilibrium first order phase transition, the system
supercools/superheats and, if driven sufficiently far from
equilibrium, it becomes unstable due to the convex anomaly in the
thermodynamic pressure. In other words, in the coexistence region
there is a range of densities and temperatures, bounded by the
spinodal lines, where the spatially uniform system is mechanically
unstable. The location of the spinodal lines is determined by the
conditions

\begin{equation}\label{eq5}
\left( \frac{\partial P}{\partial V} \right)_T=0~~~~~{\rm and} ~~~~~~
\left( \frac{\partial P}{\partial V} \right)_S=0\,,
\end{equation}
for the isothermal and isentropic spinodal lines, respectively.
Both these lines are shown in Fig. 1. From the thermodynamic
relation
\begin{align}\label{eq6}
& & \left( \frac{\partial P}{\partial V} \right)_T = \left( \frac{\partial P}{\partial V}
\right)_S {}+ \frac{T}{C_V}\left[ \left( \frac{\partial P}{\partial T} \right)_V
\right]^2\,,
\end{align}
it is clear that the isentropic spinodal lines are located inside
the instability region bounded by the isothermal spinodals and that
the two sets of lines coincide at $T=0$. In the mean field
approximation the isothermal instabilities disappear at the CEP,
where the first order transition ends, while the isentropic
spinodal curves join below the CEP. We note that the isentropic
spinodal lines will probably be modified considerably when
fluctuations are properly included. This is because the specific
heat $C_V$ diverges at the CEP \cite{pd}, while in the mean-field
approximation it remains finite. It then follows from $(\ref{eq6})$
that both the isothermal and the isentropic pressure derivatives in
$(\ref{eq5})$ vanish at the CEP. Consequently,
in a more complete description, the isentropic spinodal lines are
expected to move up in temperature and to also join at the CEP.

Since the isothermal spinodal lines join at the CEP, as shown in
Fig. 1, it is natural to explore how the charge fluctuations
develop when going beyond the critical end point into the first
order phase transition. In Fig. 2 we show the evolution of the net
quark number fluctuations along a path of fixed $T=50$ MeV  in the
$(T,n_q)$--plane. When entering the coexistence region, there is a
singularity in $\chi_{\mu\mu}$ that appears when crossing the
isothermal spinodal lines, where the fluctuations diverge and the
susceptibility changes sign. Between the spinodal lines, the
susceptibility is negative. This implies an instability of the
baryon number fluctuations when crossing the transition
between the chirally symmetric and broken phases.

\begin{figure}
\begin{center}
\includegraphics[width=8.6cm]{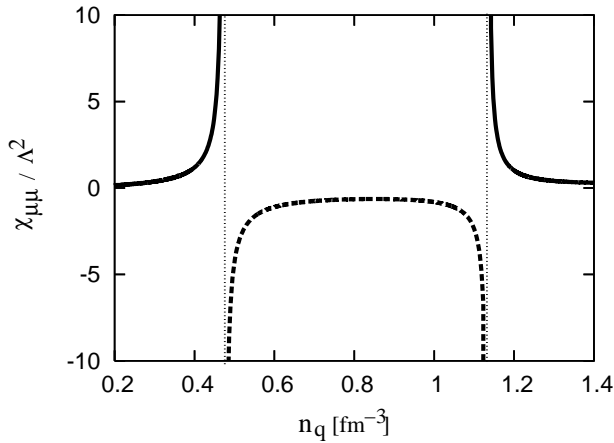}
\caption{\label{fig:Mu}  The  net quark number density fluctuations
$\chi_{\mu\mu}/\Lambda^2$ as a function of the quark  number
density $n_q$ across the first order phase transition. The
susceptibility $\chi_{\mu\mu}$ was computed in the NJL model along
the line of constant temperature $T=50$ MeV.
 }
\end{center}
\end{figure}

\begin{figure}
\begin{center}
\includegraphics[width=8.5cm, height=5.9cm]{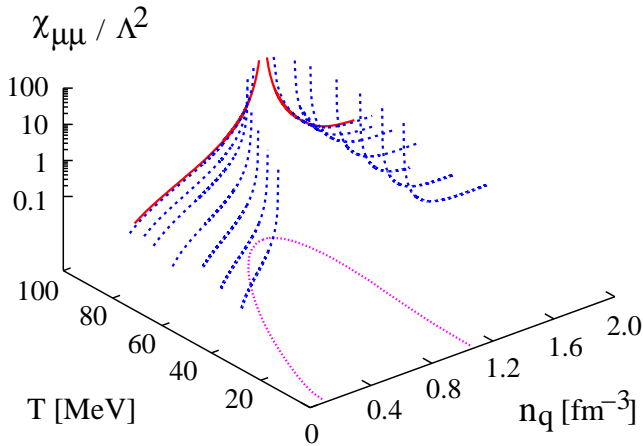}
\caption{The  net quark number susceptibility
$\chi_{\mu\mu}/\Lambda^2$ in the stable and meta stable regions, as
a function of the quark number density $n_q$ and temperature T. The
full line corresponds to the critical temperature $T_{CEP}$, and
the dashed lines to the temperatures 30, 40, 50, 60, 70, 75 and 80
MeV. The dotted line shows the projection of the isothermal
spinodal onto the $(T,n_q)$ plane.}
\end{center}
\end{figure}

An estimate of the growth rate for unstable density fluctuations,
assuming a typical wave vector $q = 100-300$ MeV and using the results
of~\cite{heiselberg2} for the collisionless regime, yields a
characteristic growth time $\tau\sim 1-10$ fm in agreement
with~\cite{gavin:hep}. We note that the detailed calculations of
ref.~\cite{heiselberg2} show that the growth rates of the most
unstable mode deep inside the spinodal region depends only weakly
on the collision rate. Thus, our estimate is expected to be valid
also in the intermediate and hydrodynamic regimes. Within the growth
time an initial fluctuation grows by a factor $e$. Since this time is
smaller than or on the order of the typical expansion time scale in
relativistic heavy-ion collisions, we conclude that the spinodal
instability may lead to observable fluctuations. A more quantitative
estimate of the expected fluctuations in heavy-ion collisions would
require a kinetic calculation, e.g. along the lines of
ref.~\cite{berdnikov}.


The behavior of $\chi_{\mu\mu}$ seen in Fig. 2 is a direct
consequence of the thermodynamic relations
\begin{align}\label{eq7}
& \left( \frac{\partial P}{\partial V} \right)_T = -
\frac{n_q^2}{V}\frac{1}{\chi_{\mu\mu}}\,,
\\
&  \left( \frac{\partial P}{\partial V} \right)_S = - \frac{n_q^2}{V}\frac{\chi_{TT} -
\frac{2 s}{n_q}\chi_{\mu T} {}+ \left( \frac{s}{n_q} \right)^2 \chi_{\mu\mu}}
{\chi_{\mu\mu}\chi_{TT} - \chi_{\mu T}^2}\,,\label{eq8}
\end{align}
which connect the pressure derivatives with the susceptibilities
$\chi_{xy}=-\partial^2\Omega /\partial x\partial y$. Along the
isothermal spinodal lines the pressure derivative in (\ref{eq7})
vanishes. Thus, for non-vanishing density $n_q$, $\chi_{\mu\mu}$
must diverge to satisfy (\ref{eq7}). Furthermore, since the
pressure derivative ${\partial P}/{\partial V}|_T$ changes sign
when crossing the spinodal line, there must be a corresponding sign
change in $\chi_{\mu\mu}$, as seen in Fig. 2. Due to the linear
relation between $\chi_{\mu\mu}$, the isovector susceptibility
$\chi_I$ and the charge susceptibility $\chi_Q$~\cite{fn1}, the
charge fluctuations are also divergent at the isothermal spinodal
line. Thus, in heavy-ion collisions, fluctuations of the baryon
number and electric charge could show enhanced fluctuations, as a
signal of the spinodal decomposition. The spinodal phase separation
can also lead to fluctuations in strangeness \cite{rans} and
isospin densities \cite{fn2}.

At the isentropic spinodal line the baryon number susceptibility is
in general finite. This is also true for the other susceptibilities
appearing in Eq. (\ref{eq8}). The isentropic spinodal line
(\ref{eq5}) corresponds to a zero of the numerator in (\ref{eq8})
and of the velocity of hydrodynamic sound waves. In the
hydrodynamic limit the instability sets in at the isothermal
spinodal, but the growth rate becomes large only at the isentropic
spinodal line~\cite{heiselberg2}.

In the case of an equilibrium first order phase transition, the
density fluctuations do not diverge; in the coexistence region the
susceptibility is a linear combination of the positive
susceptibilities above and below the phase boundary. Thus, the
fluctuations increase as one approaches the CEP along the first
order transition and decrease again in the cross over region. This
led to the prediction of a non-monotonous behavior of the
fluctuations with increasing beam energy as a signal for the
existence of a CEP~\cite{stephanov,sfr}. We stress that strictly
speaking this is relevant only for the idealized situation where
the first order phase transition takes place in equilibrium. In the
more realistic non-equilibrium system one expects fluctuations in a
larger region of the phase diagram, i.e., over a broader range of
beam energies, due to the spinodal instabilities.

The critical exponent at the isothermal spinodal line is found
to be $\gamma=1/2$, with $\chi_{\mu\mu}\sim (\mu-\mu_c)^{-\gamma}$,
while $\gamma=2/3$ at the CEP, in agreement with the mean-field
results~\cite{hatta,bj}. Thus, the singularities at the two spinodal
lines conspire to yield a somewhat stronger divergence as they join
at the CEP.
The exponents are renormalized by fluctuations, but the smooth
evolution of the singularity from the spinodal lines to the CEP,
illustrated in Fig. 3, is expected to be generic.


We have shown that the net quark number fluctuations  diverge at the isothermal spinodal
lines of the first order chiral phase transition \cite{34}. As the system crosses this
line, it becomes unstable with respect to spinodal decomposition. The unstable region is
in principle reachable in non-equilibrium systems, created e.g. in heavy ion collisions.
This means that large fluctuations of the density is expected not only at the second
order critical end point but also at a non-equilibrium first order transition. In fact,
the signal from the first-order transition may be much stronger than that from the CEP.
Model calculations suggest that the critical region of enhanced susceptibility around the
CEP is fairly small~\cite{hatta,bj,sfr}, while it is large in the spinodal region, where
the fluctuations appear due to the divergence of $\chi_{\mu\mu}$ and due to the
mechanical instability of the system. We stress that there is a close relation between
the singularities at the CEP and the spinodal lines.

The properties of different susceptibilities were obtained within
the NJL model in the mean field approximation. However, the
singular properties of charge susceptibilities in the presence of
spinodal instabilities are quite general. They appear due to the
straightforward thermodynamic relation between the pressure
derivatives and charge susceptibilities.
Thus, although the NJL model is non-confining, the results
presented in this paper are expected to be robust on a qualitative
level.

\vspace*{-0.1cm}

\section*{ Acknowledgments}

We acknowledge stimulating  discussions with  F. Karsch, V. Koch, J. Randrup,
M. Stephanov and V. Toneev. 
We thank M. Stephanov for an illuminating discussion of the singularities at the CEP.
The work of B.F. and C.S. was supported in part by the Virtual Institute of the
Helmholtz Association under the grant No. VH-VI-041. K.R. acknowledges partial support of
the Gesellschaft f\"ur Schwerionenforschung (GSI), the Polish Ministry of National
Education (MEN) and DFG under the Mercator program.



\begin{thebibliography}{99}
\bibitem{fn2} In heavy ion collisions, the rapid expansion
and the finite volume of the system suppress divergent fluctuations
of conserved charges. Consequently, the results may change
significantly, when these effects are properly accounted for. See
refs.~\cite{pd,berdnikov}
\end{thebibliography}
\end{document}